\documentstyle[aaspp]{article}
\begin{document}
\lefthead{Peletier \& Balcells}
\righthead{Surface Photometry of Bulges and Disks - Data}

\def\rpcomm#1{{\bf COMMENT by RP: #1} \message{#1}}
\def\square{\vrule height 4.5pt width 4pt depth -0.5pt}
\def\threesquares{\square~\square~\square\ }
\def\remark#1{{\threesquares\tt#1~\threesquares}}
\def\ie{{\it i.e.}}
\def\eg{{\it e.g.}}
\def\deg{\ifmmode^\circ\else$^\circ$\fi}
\def\Msun{M_\odot}
\def\etal{{\it et al.~}}
\def\deg{$^{\rm o~}$}
\def\bck{\hskip-0.35em}
\def\min#1{\ifmmode  {^{\prime}}                           
            \else    {$^{\prime}$}\fi
            \ifcat,#1{\bck}\else\null\fi\ #1}
\def\deg{\ifmmode {^{\rm o}}              
         \else {$^{\rm o}$}\fi}
\def\sec{\ifmmode {^{\prime\prime}~}       
         \else {$^{\prime\prime}~$}\fi}
\def\me{$^{-1}$}              

\onecolumn
\title{Near-Infrared Surface Photometry of Bulges and Disks
of Spiral Galaxies. The Data}

\author{R. F. Peletier{$^{1,2}$}}
\author{M. Balcells{$^{2,1}$}}

\affil{$^{1}$Kapteyn Astronomical Institute,
	Postbus 800, 9700~AV~~Groningen, Netherlands}
\affil{$^{2}$Instituto de Astrof\'\i sica de Canarias,E-38200 La Laguna,
Tenerife, Spain}

\begin{abstract}
We present optical and near-infrared (NIR) 
surface brightness and colour profiles,
in bands ranging from $U$ to $K$, 
for the disk and bulge components of a complete sample of 30 nearby 
S0 to Sbc galaxies with inclinations larger than 50$^{\rm o}$.  
We describe in detail the observations and
the determination of colour parameters. 
Calibrated monochromatic and
real-colour images are presented, as well as colour index maps.
This data set, tailored for the study of the population
characteristics of galaxy bulges,
provides useful information on the colours of inner disks as well. 
In related papers, we have used them to quantify colour gradients in bulges,
and age differentials between bulge and inner disk.
\end{abstract}


\section{Introduction}

In a recent series of papers we analyze various structural 
and population parameters of the bulge component of early-to-intermediate
type disk galaxies.  
We did this by obtaining and analyzing accurate 
surface photometry of a complete sample of inclined spiral galaxies.
In Balcells \& Peletier (\cite{BP94}, hereafter BP94) 
we derive optical colour profiles
and compare mean colours and colour gradients of bulges
to those of ellipticals.  
We find that bulges have similar or bluer colours 
than ellipticals of the same luminosity,
implying that many bulges contain younger populations, or have
lower metallicity, than corresponding ellipticals.  
In Peletier \& Balcells (\cite{PB96a}, hereafter PB96a) 
we compare the colours of bulges to those of inner disks. 
These two colours not only scale with each other,
a result already shown in BP94; colours of bulges and inner disks
are almost identical;
the bulge does not appear as a morphological component in colour index maps. 
Implied age differences are very small, 
with d$\log(age)$ ranging from 0 to 0.15.
This result, derived from NIR as well as optical data, 
is not affected by uncertainties due to dust extinction in the disk.
Finally, in Andredakis, Peletier \& Balcells (\cite{APB95}, hereafter APB95)
we develop a 2D bulge-disk decomposition,
an extension of Kent's (\cite{Kent84}) method, and show that 
surface brightness profiles of bulges 
follows an $r^{1/n}$ law, with $n$ ranging from 4-6 for S0's 
to 1 for Sc's.  
In a future paper (Peletier \& Balcells, in preparation ) we will
discuss optical-infrared colour gradients in bulges and disks; some of these
results have already been presented in Peletier \& Balcells
(\cite{PB96b}, hereafter PB96b).

The validity of these results rests first on the quality of the raw data, 
and second on the attention given to the problem of dust extinction.  
The latter is described in detail 
in the previous papers in this series.  
As for the data, 
we were fortunate that photometric conditions pervailed 
during both the optical and NIR observations;
flat-fielding, critical for sky subtraction,
was accurate to about 0.2\% of the sky,
and a factor 10 better in the near-infrared.
The sample is complete for the specified galaxy types and 
diameter and magnitude ranges.  

Optical and NIR colours of nearby galaxies contain 
important clues on the star formation history of spiral galaxies. 
Not only can one learn about the stellar contents
of these objects, but also, by comparing the properties with galaxies
at large redshift one can study the evolution of stellar populations.
In the last few years several studies of the colours of spiral galaxies
have appeared. Terndrup \etal (\cite{Terndrup++}) 
observed 43 galaxies in $J$ and $K$,
which was complemented with $r$-band photometry of 
Kent (\cite{Kent84}, \cite{Kent86}, \cite{Kent87}).
De Jong \& van der Kruit (\cite{deJ+vdK}) and de Jong (\cite{deJ}) 
published $B$, $V$, $R$, $I$ and $K$-band
photometry of 86 spiral galaxies. Both samples contain nearby spirals
of types S0 - Sc, while de Jong's sample only contains face-on or
nearly face-on galaxies. Peletier \etal (\cite{P++94}) imaged 37 galaxies in the
$K$-band, and combined this with $B$ and $R$-band images of the
ESO-LV catalog (Lauberts \& Valentijn \cite{ESO-LV}). Their sample only contains
galaxies of types Sb-Scd, at somewhat larger distances and smaller,
and which are likely to be severely affected by extinction by dust.
The current sample is complementary in type to it, and is
complementary to de Jong's sample in inclination.

The fact that our sample is biased towards early-type spiral
galaxies, with large bulges, the high inclination 
on the sky, the good sampling, the relative novelty of the  near-infrared
data, and especially the fact that we have
good-quality surface photometry in 6 bands, makes
this data set ideal to study the colour properties of bulges and inner disks
of early-type spiral galaxies. Using the opportunities of a new electronic
medium we present here all of the available data in graphical and tabular form.
In \S~\ref{observations} we describe the sample
and  the optical and NIR observations. 
In \S~\ref{reduction} we describe the complete process of 
obtaining optical and NIR colour profiles.
In \S~\ref{comparisons} we compare our results 
to NIR aperture photometry and to published NIR colour profiles.  
A discussion of the data quality is given in \S~\ref{discussion}.  

\section{Observations}
\label{observations}

\subsection{The Sample}

Our initial goal, that of obtaining colour information on bulges,
made us select galaxies at high inclinations ($i >$ 50\deg),
in order to minimize the the amount of disk light
which projects onto the bulge,  
and  to ensure that extinction from dust in the disk
does not dominate the colours measured on the bulge.  
We chose early-to-intermediate type spiral galaxies.
Dust extinction is important at high inclinations, 
but it is confined to the side of the galaxy 
were the disk is seen in front of the bulge, 
leaving the other side largely unobscured.  
We started
with a sample of 45 galaxies, comprising all galaxies with right
ascension between 13$^h$ and 24$^h$, declination above --2\deg,
galaxy type earlier than Sc, excluding barred galaxies,
apparent blue magnitude brighter than 14.0,
major axis diameter larger than 2 arcmin, absolute galactic latitude
larger than 20\deg\ and axis ratio in B larger than 1.56. 
Subsequent examination of the individual images
during the run led to the exclusion of 12 objects:
NGC 5935, 5930, 6207, 6585, 6796, 7177, 7286, 7428, 7463, 7753,
and UGC 9483, 10713,
due to being barred, very patchy all the way to the center, irregular,
or belonging to interacting systems (see BP94). NGC~7331 was not considered
due to its large size. Finally, two of the remaining 32 galaxies were too
far North to be observed from UKIRT: NGC 5308 and NGC 6361. 
All galaxies except NGC~5675 were observed in $U$, $B$, $R$, $I$ 
and $K$; 20 out of 30 were observed in $J$.
The sample is given in Table~\ref{sample}. 

\begin{table}
\begin{center}
\caption{
	\label{sample}
	\sc The Sample}
\begin{tabular}{rrrrrrrcr}	\hline\hline
UGC & NGC & Type & $\epsilon$ & M$_R^{tot}$ &  M$_R^{bul}$ & $B/D$ & $J$ & PA \\
(1) & (2) &  (3) &(4)&  (5)        &    (6)       &  (7)  & (8) & (9)\\
\hline
  8764 &  5326 &  1 &  0.50 &--22.16 & --21.22 &  0.73 & Y & -138 \\ 
  8835 &  5362 &  3 &  0.62 &--20.92 & 
	\multicolumn{1}{c}{--} & 0.00 & Y & 2 \\ 
  8866 &  5389 &  0 &  0.80 &--21.32 & --20.70 &  1.30 & Y & 93 \\
  8935 &  5422 & --2 & 0.80 &--21.96 & --21.33 &  1.28 & Y & 64 \\
  8958 &  5443 &  3 &  0.68 &--21.53 & --19.09 &  0.12 & Y & -52 \\
  9016 &  5475 &  0 &  0.68 &--20.98 & --18.72 &  0.14 & Y & -101 \\
  9187 &  5577 &  4 &  0.72 &--20.51 & --17.34 &  0.06 & Y & -38 \\
  9202 &  5587 &  0 &  0.70 &--21.07 & --19.43 &  0.29 & N & 73 \\
  9357 &  5675 &  0 &  0.68 &--22.17 & 
	\multicolumn{1}{c}{--} &  0.00 & N & 47 \\
  9361 &IC1029 &  3 & 0.76 &--21.92 & --20.66 &0.46 & N & -119 \\
  9399 &  5689 &  0 &  0.75 &--22.29 & --21.47 &  0.89 & Y & -5 \\
  9428 &  5707 &  2 & 0.75 &--21.43 & --20.11 &0.42 & N  & -59 \\
  9462 &  5719 &  2 & 0.64 &--21.60 & --20.00 &0.30 & N & 5 \\
  9499 &  5746 &  3 &  0.84 &--22.68 & --21.70 &  0.68 & Y & -99 \\
  9692 &  5838 & --3 & 0.65 &--21.89 & --20.93 &  0.71 & Y & 134 \\
  9723 &  5866 & --1 & 0.60 &--21.71 & --19.80 &  0.21 & Y & -144 \\
  9726 &  5854 & --1 & 0.70 &--21.43 & --20.18 &  0.46 & Y & -33 \\
  9753 &  5879 &  4 &  0.62 &--20.30 & --19.08 &  0.48 & Y & 89 \\
  9805 &  5908 &  3 & 0.78 &--22.88 & --22.04 &0.85 & N & 61 \\
  9914 &  5965 &  3 & 0.84 &--22.92 & --21.76 &0.53  & Y & -38 \\
  9971 &  5987 &  3 &  0.60 &--22.96 & --21.54 &  0.37 & Y & -29 \\
 10081 &  6010 &  0 &  0.75 &--21.57 & --19.88 &  0.27 & Y & 12 \\
 10856 &  6368 &  3 & 0.80 &--22.43 & --21.16 &0.45  & Y & -179 \\
 11053 &  6504 &  2 & 0.83 &--24.57 & --23.30 &0.45 & N & 135 \\
 11401 &  6757 &  0 & 0.60 &--24.60 & --22.64 &0.20 & N & 177 \\
 12080 &  7311 &  2 & 0.50 &--23.43 & --22.30 &0.55 & N & -75 \\
 12115 &  7332 & --2 & 0.74 &--21.86 & --20.52 &  0.41 & Y & -114 \\
 12306 &  7457 & --3 & 0.48 &--20.91 & --20.91 &  10.00 & Y & 35 \\
 12442 &  7537 &  4 & 0.66 &--21.37 & --19.53 &0.23 & Y & 170 \\
 12691 &  7711 & --2 & 0.56 &--22.63 & --22.63 &  10.00 & Y & -178 \\
 \hline
\hline
\end{tabular}
\end{center}
{\bf Notes to table~\ref{sample}:} Description of the columns:
\newline (1) and (2): the UGC and NGC numbers of the galaxies.
\newline (3): Type index $T$ from the 
	{\it Third Reference Catalogue of Bright Galaxies}
	(de Vaucouleurs et al. \cite{RC3}, hereafter RC3).
\newline (4): disk ellipticity derived from elliptical fits to our
	$R$-band images.
\newline (5) and (6): absolute magnitudes derived from our $R$-band photometry
	and the velocities listed in the RC3
	(uniform Hubble flow with H$_0$ = 50 km~s$^{-1}$~Mpc$^{-1}$).
\newline (7): bulge-to-disk ratio derived from an $R$-band 
	bulge-disk decomposition following Kent's (\cite{Kent84}) method.
\newline (8): observed in $J$ (Y) or not.
\newline (9): the position angle (N -- E) of the dust-free minor axis.
\end{table}

\subsection{Optical Observations}

Broad band CCD $U$, $B$, $R$ and $I$ images were obtained 
with the PF camera of the Isaac Newton telescope in June 1990.  
The optical observations are described in detail in BP94.
We used a coated 400$\times$590 GEC chip with pixel size of 0.549 arcsec.
Typical exposure times ranged from 1200 sec in $U$ to 200 sec in $R$ and $I$. 
Flat-fielding was better than 0.2\%.  
Linearity of the chip response was better than 1\% over the 
entire intensity range. 
Sky transparency was good and stable during the whole run. 
Absolute photometric errors were around 
0.05 mag in $R$ and $I$, and 0.10 mag in $B$ and $U$. The effective
seeing as measured on stellar images ranges between 1.0 and 1.5\sec.

\subsection{Near-Infrared Observations}

The NIR observations were taken on June 2-5, 1994 at UKIRT 
at Mauna Kea, Hawaii, using IRCAM3, a camera equipped with a 
256 $\times$ 256 InSb detector. 
A brief description of the observations is given in 
PB96a. The pixelsize, as measured on the
frames, was 0.291\sec. The detector was cosmetically very clean,
with less than 1\% bad pixels. 
Details can be found in Puxley \& Aspin (\cite{PuAs}).
Although relatively large for an infrared camera,
the field of IRCAM3 is still small to cover these large
galaxies. For each galaxy we observed a sequence of 10 frames: 6 frames of
the galaxy, with the center each time at a different place, and 4 of
the neighbouring background sky. For reduction, we first applied a
linearity correction to all the data. This correction was determined 
from a sequence of frames of the inside of the dome with a range of
exposure times. At the high intensity levels this correction reached
6\%. Following this
a median filtered sky frame
was made of the 4 sky-images. This frame was scaled and subtracted from each of
the object frames. Following this the object frames were flatfielded using a
flatfield obtained from all dark sky frames of that night. Our 
experience is that for this kind of InSb array these flatfields 
are of the same quality as dome flatfields. Finally, a mosaic
was made by combining the 6 frames, using the galaxy center as the reference
point, and adjusting the sky background of each frame individually. 
Details about how the mosaics were made can be found e.g. in 
Peletier (\cite{P93}). The final mosaics have a size of aproximately
120 $\times$ 100\arcsec. 
Since the galaxies are inclined to the line-of-sight, 
the mosaics are large enough to determine and subtract the
residual sky background for most of the galaxies.
The seeing FWHM, as measured on
bright stars in the images, ranges from 0.8 to 1.2\arcsec.
Our typical integration times per readout were 6 or 10s in $K$ and 10 or 
20s in $J$. After typically 90s the coadded frame was written to disk
and the telescope moved to a slightly different position. 

The conditions were photometric during the 2nd and 3rd night. During these
two nights we reobserved briefly all galaxies of the first night, to 
obtain photometric zero points. During each of the last two nights,
we observed 8-10 UKIRT faint standard stars of the list of Casali (unpublished), 
with at least 4 observations
per standard star in different positions on the chip. The RMS accuracy
of the standard star observations was 0.03 mag in $K$ and 0.035 mag in $J$.
The $J-K$ colour term was negligible. In the infrared 
no corrections were made for 
galactic extinction and no K-correction because of redshift of the 
spectrum was applied. These corrections however were made in the optical
(see BP94).

\section{Data reduction}
\label{reduction}
\subsection{Near-infrared data reduction}

Essential in the derivation of NIR mosaics is to derive good estimates for the
sky background. Terndrup \etal (\cite{Terndrup++}) discuss this point extensively.
The variability in the sky, on levels of 10 - 100\% in the near-infrared
in the course of one night 
and on levels of percents during the taking of the mosaic, is so large
that a residual sky background had to be determined on the final mosaic.
Fortunately, IRCAM3 is very stable,
as compared to for example 256 $\times$ 256 HgCdTe detectors, making flatfielding to better than 0.1\% easily possible. 
As a test, we divided 2 
sky-frames, taken 10 minutes apart, by each other, after having
subtracted the dark current. The non-random non-uniformities
on scales of $\sim$ 10 -- 20''  
are about 0.05\%. In our case, where we take first
three galaxy-frames, then four sky-frames, and then three galaxy-frames,
and where we first subtract the median of the sky's before flatfielding,
we were in general able to achieve an accuracy of 0.02\% of sky.
When the conditions were not photometric, i.e. during parts of
the first night, the flatfielding accuracy is somewhat worse.

Even though we use a larger format array than has been done in the past, 
and even after mosaicing, some galaxies are larger than the entire frame. 
In these cases, the sky determination is most probably an overestimate. 
We estimate by how much by taking a much larger $I$-band 
frame, assuming that the colour gradients are small in the outer parts, 
and measuring
the contribution in $I$ at about 50'' from the center on the galaxy's
minor axis. This contribution in $J$ and $K$ generally corresponds
to a surface brightness level between 23 and 25 mag/sq. arcsec in $J$ and
between 22 and 24 mag/sq. arcsec in $K$. 
Although different from galaxy to galaxy, a 1$\sigma$ error 
in the sky corresponds in $U$, $B$, $R$, $I$, $J$ and $K$ appr. to
resp 28.0, 29.0, 27.0, 26.0, 23.0 and 21.7 mag/sq. arcsec (for details 
about the numbers in the optical see BP94).

\subsection{Optical and NIR colour profiles}

After reduction we rotated and rebinned the $J$ and $K$ mosaics to 
align them with the optical images. 
The scaling and rotation was determined on some frames 
with many stars, then applied to all the NIR frames.
The frames for the various passbands were then aligned using stars.
Some NIR frames do not contain stellar images. 
In those cases we used the galaxy center 
to align the NIR images with the $I$-band frame. The center shift
between $I$ and $K$, due to dust extinction, is very small
(numbers are given in Table~\ref{ShiftsTable}, see  \S~\ref{centershifts}). 
Only for NGC~5866 (M~102) was this procedure especially difficult,
due to the fact that a very strong dustlane obscures the
center even in $K$.
Given the lack of colour structures across stellar images in the colour maps, 
we estimate that the alignment is accurate to of order 0.2''.

\subsection{Bulge colours and colour gradients}
\label{BulCol}

Paper I describes how we determine 
bulge optical colours and colour gradients.
We use the same approach for the determination of near-infrared colours
and gradients.  Briefly, we follow the following steps.

First, we determine the galaxy center as the central luminosity peak 
of the $K$-band galaxy image. 
With the exception of NGC~5866, whose prominent dust lane 
has been described in the previous section,
the $K$-band peak always lies at the center of
symmetry of the galaxy, as determined by ellipse fitting to the entire
bulge region.  The $K$-band peak is thus a suitable choice of center.  

Second, we determine the galaxy's orientation on the sky
by fitting ellipses to the $K$-band frames.
For these highly inclined galaxies 
the disk major axis orientation is well constrained 
by the ellipticity profile, which becomes flat in the outer parts.  
The orientation of each galaxy is given in Table~\ref{sample}
as the position angle of the semi-minor axis which shows 
the least amount of dust obscuration (see below).  

Once center and orientation are determined, 
we extract surface brightness profiles on both semi-minor axes. 
We use azimuthal averages on wedge-shaped apertures 
with full-width 22.5\deg.
Colour profiles are determined by subtracting calibrated profiles
from each other. 
Appendix A gives the wedge surface brightness profiles in $R$ 
and the colour profiles, in tabular and graphical form,
for the two semi-minor axes of each galaxy.

As a result of the high inclination of our sample,
colour structures associated to reddening by dust in the disk 
are commonly seen on one semi-minor axis of each galaxy.
On the opposite semi-minor axis the colour profiles are generally featureless
and scale-free, ie. linear when plotted against log($r$).  

To determine bulge central colours and colour gradients
we fit a linear relation to the colour profiles between an inner
and an outer radius, as described in BP94. We used a minimum 
inner cutoff radius of 1 seeing FWHM, to avoid any effects of seeing
and bad centering, but in cases where the colour gradients changed
significantly inside the bulge, we moved the inner cutoff out to
the radius where this change occurred. Changes like this can be 
caused by dust extinction in the nuclear area, or by a change
in stellar populations, similar to elliptical galaxies (e.g. Surma
\& Bender \cite{SurmaBender}).
We put the outer radius
at the place where, according to a bulge-disk decomposition,
50\%  of the light comes from the bulge. In some galaxies with large
bulges the outer radius was set at 30''. The inner and outer radii
are listed in Table~\ref{TabBulGrad}, where we also 
present the bulge colour gradients. The optical gradients 
presented here are not identical as those presented in BP94,
especially as a result of different centering and sometimes
different inner radii, but the differences are small, 
and give an indication of the error in these measurements. The
errors presented in Table 3 are formal errors of the fit.

To give characteristic values for the bulge and disk colours
we follow the convention we use 
in the previous papers in this series, namely
we extract the bulge colours at 0.5 r$_{\rm eff}$,
or 5'', if this was larger, and the disk colours at 2 major axis 
$K$-band scale lengths.
Table~\ref{ColTab} lists the characteristic colours of bulges and disks. 
The effective radii of the bulges in $K$ have been determined from the 
two-dimensional bulge-disk decomposition of the $K$-band frames
in APB95; they are very similar, but not identical, to those
given in BP94.  

{\scriptsize
\begin{table}
\begin{center}
\caption{
	\label{TabBulGrad}
	\sc Bulge colour gradients}
\begin{tabular}{lcccccccccccc}	\hline\hline
NGC & \multicolumn{2}{c}{Fitting Range (arcsec)} & $\nabla(B-R)$ & $\pm$ &
$\nabla(U-R)$ & $\pm$ & $\nabla(R-K)$ & $\pm$ & $\nabla(R-I)$ & $\pm$ &
$\nabla(J-K)$ & $\pm$ \\
(1) & (2) & (3) & (4) & (5) & (6) & (7) & (8) & (9) & 
(10) & (11) & (12) & (13) \\
\hline
 5326 &  1.8&   10.5&   -0.21&   0.02&   -0.33&   0.02&   -0.34&   0.04&    0.00&   0.02&   -0.19&   0.02 \\
 5362 &  1.6&    6.0&   -0.10&   0.02&   -0.35&   0.03&   -0.13&   0.04&   -0.14&   0.02&  ~   &   ~  \\
 5389 &  3.0&   14.0&   -0.18&   0.03&   -0.41&   0.06&   -0.15&   0.01&    0.02&   0.02&    0.03&   0.01 \\
 5422 &  2.0&   14.0&   -0.15&   0.02&   -0.31&   0.01&   -0.17&   0.02&   -0.04&   0.02&   -0.09&   0.01 \\
 5443 &  5.0&   10.0&   -0.13&   0.06&   -0.38&   0.07&   -0.30&   0.04&   -0.10&   0.13&    0.13&   0.06 \\
 5475 &  3.0&   11.0&   -0.10&   0.06&   -0.36&   0.05&   -0.38&   0.08&   -0.11&   0.07&   -0.18&   0.03 \\
 5577 &  1.5&   11.0&   -0.43&   0.02&   -0.85&   0.04&   -0.44&   0.05&   -0.15&   0.02&   -0.08&   0.03 \\
 5587 &  1.5&   20.0&   -0.33&   0.02&   -0.73&   0.04&   -0.15&   0.06&   -0.07&   0.02& ~    &   ~ \\
 5675 &  5.0&   10.2&   -0.22&   0.07&    ~   &   ~   &   -0.25&   0.06&   -0.02&   0.04& ~    &   ~ \\
 IC 1029 &  2.5&   10.6&   -0.45&   0.03&   -0.77&   0.10&   -0.20&   0.04&   -0.09&   0.04&  ~   &   ~ \\
 5689 &  3.0&   10.0&   -0.08&   0.01&   -0.27&   0.02&   -0.22&   0.02&   -0.04&   0.02&    0.04&   0.02 \\
 5707 &  1.6&   7.2&   -0.35&   0.02&   -0.62&   0.02&   -0.07&   0.02&   -0.06&   0.01& ~    &  ~  \\
 5719 &  1.4&   12.0&   -0.34&   0.02&   -0.61&   0.04&   -0.75&   0.02&   -0.12&   0.01& ~    &  ~  \\
 5746 &  2.5&   21.0&   -0.25&   0.01&   -0.50&   0.02&   -0.32&   0.02&   -0.12&   0.01&   -0.10&   0.01 \\
 5838 &  3.0&   20.0&   -0.10&   0.02&   -0.28&   0.02&   -0.13&   0.01&   -0.09&   0.01&   -0.02&   0.01 \\
 5854 &  1.4&   9.2&    0.11&   0.02&   -0.03&   0.01&   -0.12&   0.01&   -0.11&   0.02&    0.03&   0.01 \\
 5866 &  6.0&   15.0&   -0.15&   0.02&   -0.30&   0.03&   -0.30&   0.04&   -0.10&   0.02&   -0.07&   0.04 \\
 5879 &  1.6&   7.0&   -0.40&   0.03&   -1.00&   0.03&   -0.65&   0.02&   -0.24&   0.01&   -0.27&   0.02 \\
 5908 &  9.0&   30.0&   -0.03&   0.07&   -0.08&   0.07&    0.01&   0.10&   -0.05&   0.08&   ~  &   ~ \\
 5965 &  1.4&   11.0&   -0.17&   0.02&   -0.35&   0.04&   -0.29&   0.03&   -0.10&   0.01&    0.05&   0.01 \\
 5987 &  1.5&   13.0&    0.08&   0.01&   -0.15&   0.02&   -0.35&   0.02&   -0.24&   0.02&   -0.08&   0.02 \\
 6010 &  1.4&   6.0&   -0.26&   0.01&   -0.62&   0.02&   -0.30&   0.03&   -0.01&   0.02&   -0.04&   0.02 \\
 6368 &  1.5&   10.0&   -0.24&   0.03&   -0.28&   0.10&   -0.20&   0.01&   -0.04&   0.02&   -0.03&   0.02 \\
 6504 &  1.7&  8.1&   -0.15&   0.01&   -0.27&   0.01&   -0.01&   0.02&   -0.04&   0.02&   ~  &   ~ \\
 6757 &  1.7&   8.0&   -0.11&   0.02&    0.02&   0.03&   -0.24&   0.04&   -0.25&   0.02&   ~  &   ~ \\
 7311 &  1.6&   15.0&   -0.43&   0.03&   -1.10&   0.09&   -0.24&   0.04&   -0.22&   0.03&   ~  &   ~ \\
 7332 &  1.5&   8.2&    0.10&   0.03&   -0.18&   0.02&   -0.32&   0.03&   -0.26&   0.03&   -0.04&   0.01 \\
 7457 &  1.3&   20.0&    0.08&   0.02&    0.08&   0.02&   -0.25&   0.01&   -0.04&   0.01&   -0.06&   0.01 \\
 7537 &  2.0&   12.0&   -0.44&   0.05&   -0.49&   0.07&   -0.43&   0.06&   -0.03&   0.02&   -0.18&   0.05 \\
 7711 &  1.4&   11.5&   -0.17&   0.03&   -0.27&   0.02&   -0.26&   0.02&   -0.14&   0.02&   -0.13&   0.01 \\
\hline
\hline
\end{tabular}
\end{center}
{\bf Notes to table~\ref{TabBulGrad}:} Description of the columns:
\newline Columns (2) and (3) display the radial
range on the minor axis used to fit the bulge color gradients. Columns (4),
(6), (8), (10) and (12) show the gradient of colour against
logarithmic radius $\Delta(Colour)/\Delta(\log~r)$. In columns (5),
(7), (9), (11) and (13) the formal error of this fit is given.
\end{table}
}

\begin{table}
\begin{center}
\caption{
	\label{ColTab}
	Colours of bulges and disks
	}
\begin{tabular}{lcccccccc}
\hline
\hline
Galaxy & (B-R)$_B$ & (B-R)$_D$ & (U-R)$_B$ & (U-R)$_D$ & 
(R-K)$_B$ & (R-K)$_D$ & (J-K)$_B$ & (J-K)$_D$ \\
\hline

5326   &   1.444  &    1.405   &   1.946   &   1.845  &    2.604  &     2.56  &    0.874  &    0.911 \\
5362   &   1.144  &     0.97   &    1.46   &   1.012  &    2.414  &    2.274  &     --    &     --   \\
5389   &   1.439  &    1.434   &   1.741   &   1.772  &    2.678  &    2.662  &    0.874  &    0.921 \\
5422   &   1.462  &    1.376   &   1.901   &   1.804  &    2.704  &    2.709  &    0.867  &    0.913 \\
5443   &   1.521  &    1.354   &   2.057   &   1.776  &    2.461  &    2.485  &    0.931  &    0.925 \\
5475   &   1.474  &    1.506   &   1.876   &   1.885  &    2.498  &    2.474  &    0.867  &    0.883 \\
5577   &     1.3  &    1.061   &   1.235   &   0.821  &    2.539  &    2.171  &    0.835  &    0.589 \\
5587   &   1.471  &    1.423   &   1.914   &   1.777  &    2.564  &    2.602  &     --    &     --   \\
5675   &   1.472  &    1.329   &     --    &    --    &    2.58   &    2.458  &     --    &     --   \\
IC 1029   &   1.402  &     1.25   &   1.808   &   1.614  &    2.492  &     2.52  &     --    &     --   \\
5689   &   1.437  &    1.482   &   1.861   &   1.896  &    2.705  &    2.638  &    0.902  &    0.918 \\
5707   &    1.53  &    1.328   &   1.874   &   1.629  &    2.711  &    2.588  &     --    &     --   \\
5719   &    1.61  &    1.402   &   2.059   &   1.759  &    2.933  &    2.435  &     --    &     --   \\
5746   &   1.595  &     1.65   &   2.162   &   2.214  &    2.821  &    2.851  &    0.982  &    0.979 \\
5838   &   1.488  &    1.435   &   2.039   &   1.871  &    2.725  &    2.679  &    0.866  &    0.838 \\
5866   &   1.451  &    1.457   &   1.841   &   1.682  &    2.609  &     2.44  &    0.863  &     0.94 \\
5854   &   1.392  &    1.442   &   1.742   &    1.79  &     2.48  &    2.495  &    0.922  &    0.781 \\
5879   &   1.271  &    1.314   &   1.279   &   1.293  &    2.404  &    2.476  &    0.779  &    0.833 \\
5908   &   1.441  &    1.547   &   1.765   &   1.908  &    2.595  &    2.984  &     --    &     --   \\
5965   &   1.408  &    1.444   &   1.978   &   2.038  &    2.667  &    2.731  &    0.901  &    0.962 \\
5987   &    1.49  &    1.477   &   1.962   &   1.886  &    2.966  &    2.667  &     1.18  &    1.074 \\
6010   &   1.435  &    1.504   &   1.805   &   1.759  &    2.535  &     2.66  &    0.954  &    1.046 \\
6368   &   1.723  &    1.607   &   2.031   &    1.91  &    3.121  &    2.976  &    0.896  &    0.929 \\
6504   &   1.789  &    1.746   &   2.118   &   1.721  &    2.633  &    2.381  &     --    &     --   \\
6757   &   1.434  &     1.38   &   1.672   &    1.46  &    2.707  &     2.76  &     --    &     --   \\
7311   &   1.523  &    1.348   &   1.963   &   1.479  &    2.827  &    2.726  &     --    &     --   \\
7332   &   1.353  &    1.413   &   1.802   &   1.753  &    2.393  &    2.165  &    0.829  &    0.658 \\
7457   &   1.287  &    1.325   &   1.675   &   1.618  &    2.403  &    2.172  &    0.862  &    0.829 \\
7537   &   1.285  &    1.242   &   1.267   &   1.138  &    2.593  &    2.379  &    0.875  &    0.866 \\
7711   &   1.405  &    1.464   &   2.022   &   2.083  &    2.653  &    2.551  &    0.901  &    0.838 \\
\hline
\end{tabular}
\end{center}
{\bf Notes to table~\ref{ColTab}:} Bulge colours are given 
at 0.5 r$_{\rm eff}$ or
at 5'', whichever larger, on the minor axis. Disk colors are given at two
major axis $K$-band scale lengths, and have been measured using the 
major axis 'dustfree' wedges.
\end{table}

\subsection{Disk surface brightness, colours and scale length ratios}
\label{DisCol}

For the extraction of disk surface brightness distributions,
we center wedge apertures slightly away (15\deg)
from each semi-major axis, on the least dusty side; 
major-axis apertures often cut over strong dust lanes on the disk, 
which we want to avoid.  The disk apertures have a 
full-width of 10\deg. 
Using the $K$-band center we average azimuthally
the light in the wedge. Colour profiles are determined by subtracting
calibrated profiles from each other. 
The profiles for the two wedges are finally averaged.  
The difference between the profiles for the two wedges
is the main source of uncertainty on the disk surface brightness profile;
the profiles do not reach the disk faint outer parts hence 
sky uncertainty does not contribute to surface brightness errors.  
We call these near-major-axis profiles the "disk profiles" 
although, clearly, bulge light contributes in the central parts.
No geometric scaling has been applied to project
the profiles to the major axis.  
Such scaling is applied later whenever needed, eg. 
for the determination of disk scale lengths.  
Colour profiles are derived from the ratios of surface brightness profiles.
We give surface brightness and colour profiles for the disk profiles
in tabular and graphical form in Appendix A.  

We next measure the colour gradients in the disk 
by fitting linear laws to the colour profiles against linear radius. 
The results from the fit are given in Table~\ref{DiskPars}.
The gradients are
presented per projected $K$-band scale length, also given
in Table~\ref{DiskPars}.
The radial range is chosen in such a way that we fit the exponential 
colour gradients outside the region of the bulge, and that a radial region
is as large as possible and present in all passbands.
In any case, we don't see strong variations in slope of colour as
a function of radius when going from the bulge to the disk 
(see Appendix A and PB96a).
The fitting range is the same for each passband. 

{\scriptsize
\begin{table}
\begin{center}
\caption{
	\label{DiskPars}
	\sc Disk parameters}
\begin{tabular}{lccrccccccccccc}	\hline\hline
NGC & \multicolumn{2}{c}{Fitting Range} & h$_K$ & $\pm$ & (B-R)$_0$ & 
${\Delta(B-R)}\over{\Delta(h_K)}$ & (U-R)$_0$ & ${\Delta(U-R)}\over
{\Delta(h_K)}$ &
(R-K)$_0$ & ${\Delta(R-K)}\over{\Delta(h_K)}$ & 
(R-I)$_0$ & ${\Delta(R-I)}\over{\Delta(h_K)}$ &
(J-K)$_0$ & ${\Delta(J-K)}\over{\Delta(h_K)}$ \\
(1) & (2) & (3) & (4) & (5) & (6) & (7) & (8) 7 (9) &
(10) & (11) & (12) & (13) & (14) & (15) \\
\hline
 5326   &  15 & 45& 9.1  &    0.1&   1.61 & -0.05&   2.23&  -0.12&  3.28 &-0.22&  0.66&  0.02&  0.88&  0.01  \\
 5362   &  10 & 40& 9.2  &    0.1&   1.24 & -0.08&   1.55&  -0.18&  2.63 &-0.16&  0.60& -0.05&   ~  &   ~\\
 5389   &  15 & 40& 17.9 &    1.9&   1.79 & -0.04&   2.11&   0.00&  3.02 & 0.05&  0.93& -0.05&  1.00& -0.05\\
 5422   &  10 & 45& 10.9 &    0.1&   1.66 & -0.06&   2.20&  -0.08&  2.71 &-0.01&  0.70&  0.00&  0.84&  0.01  \\
 5443   &  10 & 45& 17.6 &    0.3&   1.55 & -0.10&   2.18&  -0.31&  2.64 &-0.09&  0.61&  0.04&  0.93&  0.07  \\
 5475   &  10 & 40& 9.0  &    0.1&   1.53 & -0.04&   1.94&  -0.03&  2.50 &-0.03&  0.65& -0.03&  0.84& -0.01  \\
 5577   &  10 & 45& 20.5 &    0.1&   1.42 & -0.18&   1.28&  -0.14&  2.80 &-0.19&  0.61& -0.01&  1.00& -0.12  \\
 5587   &  15 & 45& 8.4  &    0.1&   1.71 & -0.10&   2.25&  -0.20&  3.02 &-0.12&  0.74& -0.07&   ~  &   ~\\
 IC 1029&  10 & 35& 9.6  &    0.1&   1.98 & -0.31&   2.48&  -0.38&  3.45 &-0.23&  0.84& -0.07&   ~  &   ~\\
 5675   &  10 & 45& 11.3 &    0.4&   1.75 & -0.17&    ~  &    ~  &  2.92 &-0.22&  1.00& -0.18&  2.64& -0.30\\  
 5689   &  15 & 45& 11.4 &    0.3&   1.85 & -0.12&   2.45&  -0.28&  3.16 &-0.29&  0.79& -0.06&  1.04& -0.06  \\
 5707   &  10 & 45& 12.6 &    0.1&   1.67 & -0.15&   2.10&  -0.17&  2.74 & 0.01&  0.75& -0.05&   ~  &   ~\\
 5719   &  15 & 45& 10.7 &    0.4&   2.29 & -0.21&   2.95&  -0.28&  3.57 &-0.20&  0.85& -0.07&   ~  &   ~\\
 5746   &  10 & 35& 16.4 &    0.8&   2.29 & -0.24&   3.14&  -0.57&  3.01 & 0.71&  0.78&  0.11&  0.83&  0.31\\  
 5838   &  15 & 45& 17.9 &    0.2&   1.50 & -0.01&   2.22&  -0.11&  2.83 &-0.09&  0.66& -0.01&  0.88&  0.01  \\
 5854   &  10 & 45& 10.5 &    0.1&   1.48 &  0.01&   1.49&   0.14&  2.45 & 0.00&  0.60& -0.01&  0.92&  0.01  \\
 5866   &  15 & 45& 12.2 &    0.5&   1.62 & -0.08&   1.96&  -0.09&  2.88 &-0.14&  0.65& -0.01&  0.94& -0.08  \\
 5879   &  10 & 30& 8.4  &    0.1&   1.59 & -0.13&   1.45&  -0.13&  3.26 &-0.31&  0.65& -0.04&  1.12& -0.13  \\
 5908   &  10 & 45& 10.2 &    0.1&   1.89 & -0.13&   1.64&  -0.06&  4.94 &-0.53&  1.19& -0.12&   ~  &   ~\\
 5965   &  10 & 45& 12.2 &    0.4&   1.98 & -0.23&   3.14&  -0.58&  3.49 &-0.17&  0.92& -0.10&  0.95&  0.13\\  
 5987   &  10 & 45& 12.6 &    0.3&   1.95 & -0.04&   2.79&  -0.15&  3.95 &-0.29&  0.89& -0.05&  1.30&  0.01  \\
 6010   &  10 & 45& 11.5 &    0.2&   1.55 &  0.00&   2.21&  -0.18&  2.56 & 0.04&  0.66& -0.03&  0.80&  0.15  \\
 6368   &  10 & 40& 13.2 &    0.3&   1.93 & -0.20&   2.48&  -0.38&  3.34 &-0.16&  0.87& -0.09&  0.86&  0.08  \\
 6504   &  15 & 45& 14.4 &    0.2&   1.92 & -0.11&   2.51&  -0.24&  2.87 &-0.09&  0.77& -0.08&   ~  &   ~\\
 6757   &  10 & 45& 9.9  &    0.1&   1.57 & -0.06&   1.92&  -0.04&  2.73 & 0.01&  0.57&  0.00&   ~  &   ~\\
 7311   &  10 & 40& 9.0  &    0.1&   1.73 & -0.08&   2.47&  -0.25&  3.12 &-0.14&  0.81& -0.05&   ~  &   ~\\
 7332   &  10 & 35& 11.4 &    0.3&   1.46 & -0.06&   1.96&  -0.08&  2.46 &-0.04&  0.52& -0.01&  0.95& -0.12\\  
 7457   &  10 & 45& 16.0 &    0.1&   1.36 & -0.04&   1.79&  -0.06&  2.47 &-0.11&  0.57& -0.01&  0.89& -0.06  \\
 7537   &  15 & 30& 8.8  &    0.6&   1.43 & -0.11&   1.60&  -0.25&  2.97 &-0.25&  0.85& -0.17&  1.27& -0.11  \\
 7711   &  10 & 40& 10.1 &    0.1&   1.54 & -0.05&   2.12&  -0.08&  2.67 & 0.00&  0.58&  0.02&  0.91&  0.02  \\
\hline
\end{tabular}
\end{center}
{\bf Notes to table~\ref{DiskPars}:} In columns (2) and (3) the fitting
range along the wedge is given in arcsec. Columns (4) and (5) give 
the $K$-band scale length in arcsec, calculated using the average of the 2 wedges at 15\deg\ from the
major axis on the 'dustfree' side, and its formal
fit error. To obtain the major axis scale length, multiply this
number by 1.03.
In columns (6), (8), (10), (12) and (14) are given the 
extrapolated central colors of the disk, as fitted. Columns (7),
(9), (11), (13) and (15) contain the color gradients, 
along a wedge aperture at 15\deg\ from the major axis, per $K$-band
scale length.
\end{table}
}

\begin{figure}
\plotone{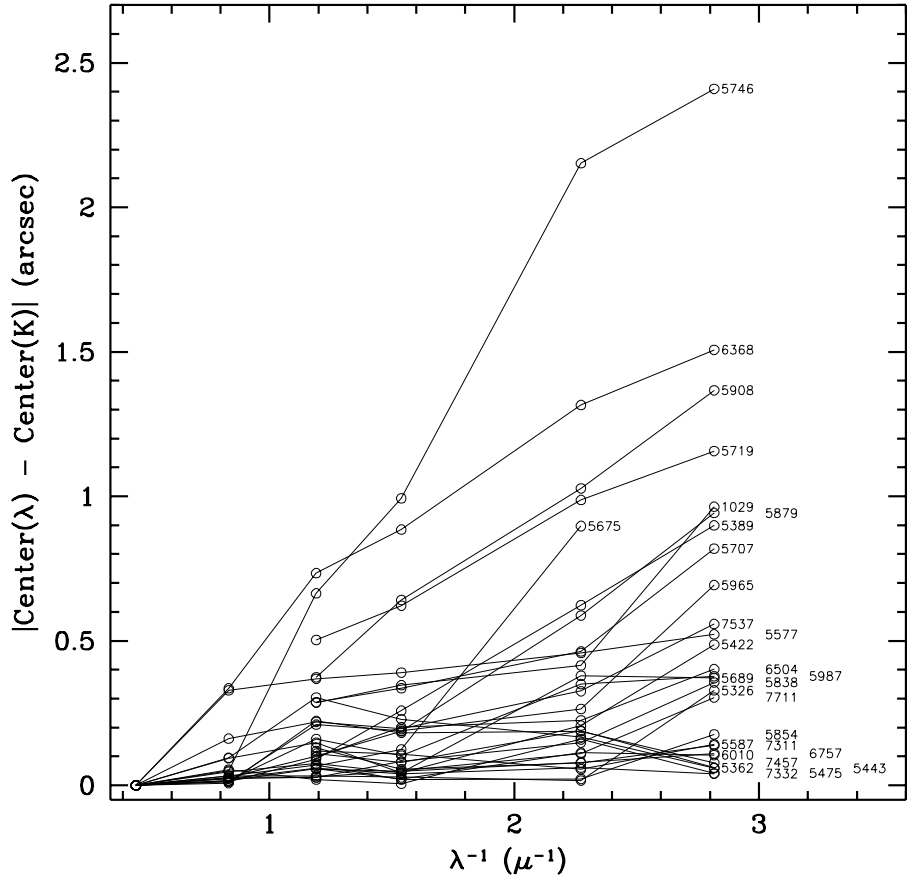}
\caption{\label{shifts}
	Shift of the position of the galaxy center as compared to the $K$-band,
plotted against reciprocal wavelength.}
\end{figure}

\subsection{Center shifts}
\label{centershifts}

A useful indicator for the presence of dust in inclined galaxies 
is the change in the position of the center as a function of
wavelength. This shift occurs when a dustlane
in front of the center of light obscures more light on one side
of the center than on the other, causing the observed center to shift
as a function of optical depth, or passband. 
In Table~\ref{ShiftsTable} we give
the shift of the peak in arcsec between the $K$-band and the other
bands. Shifts are approximately along the minor axis.
Typical errors, including the uncertainty of the determination 
of the luminosity peaks and the errors of alignment, 
are 0.2--0.3 arcsec. 
We plot the shifts as a function of reciprocal wavelength
in Fig.~\ref{shifts}, showing that they increase with $\mu^{-1}$
in an approximately linear way.

\begin{table}
\begin{center}
\caption{
	\label{ShiftsTable}
	\sc Shifts of the galaxy luminosity peaks with respect to the 
	$K$-band position}
\begin{tabular}{rccccc}	\hline\hline
NGC & $\Delta$(JK) (arcsec) & $\Delta$(IK) (arcsec)
& $\Delta$(RK) (arcsec) & $\Delta$(BK) (arcsec)  & $\Delta$(UK) (arcsec) \\
\hline
5326	 &   0.032  &   0.033  &   0.028 &    0.017   &  0.328 \\
5362	 &    --    &   0.109  &   0.080 &    0.188   &  0.061 \\
5389	 &   0.012  &   0.090  &   0.258 &    0.623   &  0.899 \\
5422	 &   0.047  &   0.075  &   0.043 &    0.206   &  0.487 \\
5443	 &   0.095  &   0.304  &   0.229 &    0.167   &  0.059 \\
5475	 &   0.054  &   0.020  &   0.005 &    0.159   &  0.042 \\
5577	 &   0.328  &   0.369  &   0.390 &    0.458   &  0.522 \\
5587	 &    --    &   0.118  &   0.108 &    0.055   &  0.142 \\
5675	 &    --    &   0.066  &   0.124 &    0.897   &   -- \\
IC 1029	 &    --    &   0.286  &   0.346 &    0.415   &  0.964 \\
5689	 &   0.094  &   0.148  &   0.044 &    0.379   &  0.371 \\
5707	 &    --    &   0.288  &   0.336 &    0.463   &  0.819 \\
5719	 &    --    &   0.503  &   0.622 &    0.987   &  1.157 \\
5746	 &   0.050  &   0.664  &   0.993 &    2.152   &  2.410 \\
5838	 &   0.025  &   0.027  &   0.082 &    0.148   &  0.356 \\
5866	&    --    &    --    &    --   &     --     &   -- \\
5854	 &   0.032  &   0.070  &   0.020 &    0.022   &  0.176 \\
5879	 &   0.027  &   0.101  &   0.187 &    0.588   &  0.942 \\
5908	 &    --    &   0.374  &   0.641 &    1.028   &  1.367 \\
5965	 &   0.012  &   0.210  &   0.188 &    0.264   &  0.693 \\
5987	 &   0.016  &   0.160  &   0.103 &    0.350   &  0.378 \\
6010	 &   0.017  &   0.057  &   0.022 &    0.113   &  0.105 \\
6368	 &   0.335  &   0.734  &   0.885 &    1.316   &  1.507 \\
6504	 &    --    &   0.097  &   0.201 &    0.225   &  0.402 \\
6757	 &    --    &   0.056  &   0.049 &    0.079   &  0.110 \\
7311	 &    --    &   0.130  &   0.056 &    0.077   &  0.138 \\
7332	 &   0.023  &   0.059  &   0.040 &    0.061   &  0.040 \\
7457	 &   0.008  &   0.222  &   0.182 &    0.187   &  0.078 \\
7537	 &   0.161  &   0.220  &   0.195 &    0.326   &  0.558 \\
7711	 &   0.023  &   0.028  &   0.050 &    0.109   &  0.304 \\
\hline
\end{tabular}
\end{center}
\end{table}

\subsection{Comparison with other authors}
\label{comparisons}

In BP94 we compared our optical data with various sources from
the literature. Here we compare the infrared data. 

\subsubsection{Comparison with literature surface photometry}

\begin{figure}
\plotone{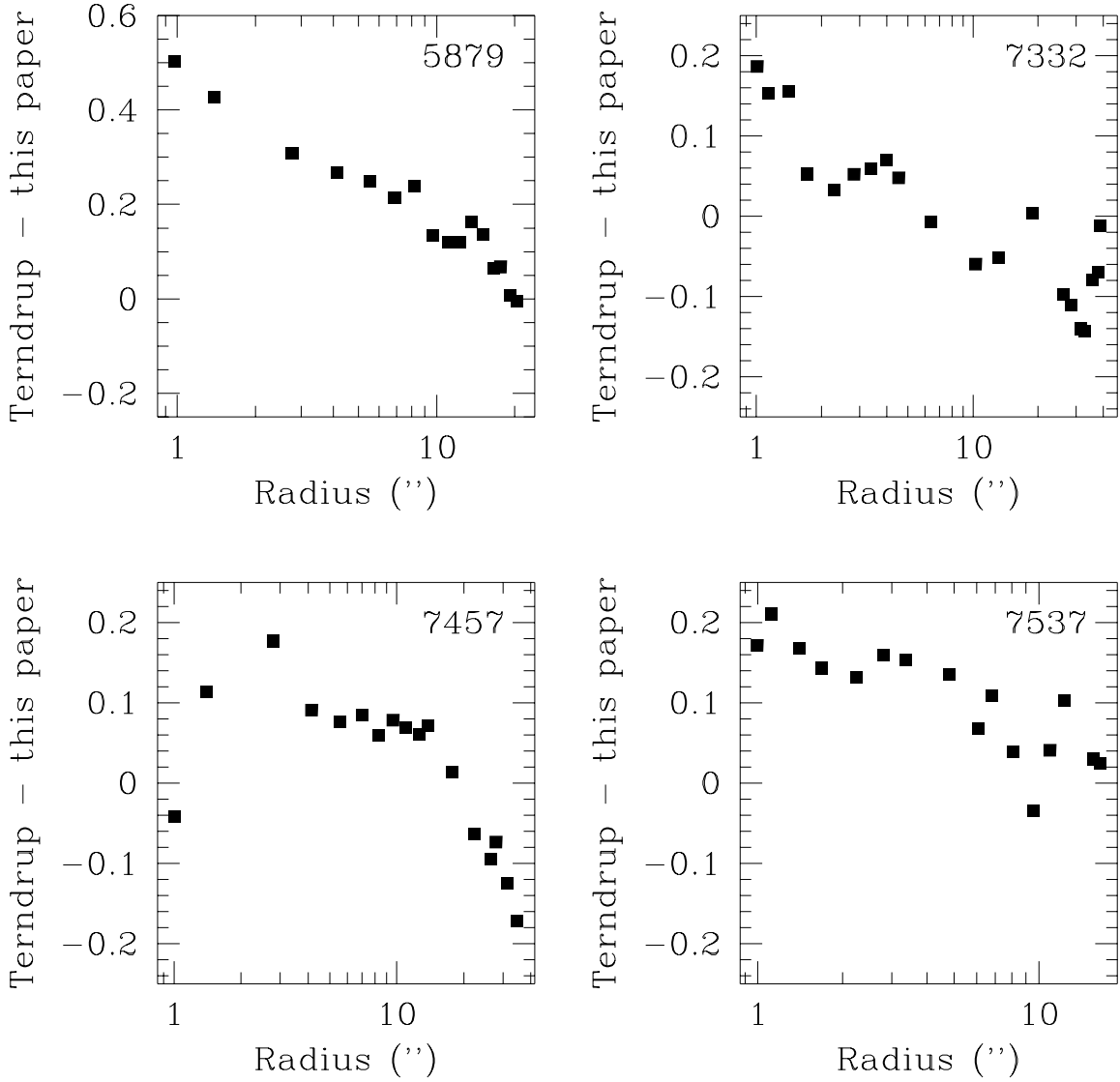}
\caption{
	\label{CompTerndrup}
	Comparison with the $K$-band major axis profiles of Terndup \etal
(1994). }
\end{figure}

Since de Jong's
sample is orientated towards face-on galaxies and ours towards edge-on,
there is no overlap between the two samples. We have 4 galaxies in common
with Terndrup \etal (\cite{Terndrup++}). For the comparison we have determined
elliptically averaged major axis profiles in the same way as was done
by Terndrup \etal (\cite{Terndrup++}) and then made the comparison 
(see Fig.~\ref{CompTerndrup}), with their corrected data (see
Terndrup \etal (\cite{Terndruperr}).
The differences near the center can be explained well by the fact
that our seeing and sampling was better than that of Terndrup
\etal (\cite{Terndrup++}). In the outer parts of NGC 7457 the difference might be 
explained by sky background errors, since their field was smaller.
However, there seems to be a general offset between the two datasets,
of about 0.05 - 0.10 mag, in the sense that the profiles of Terndrup 
\etal are fainter. This is however not the case for NGC 7332, the best
studied galaxy.

As for the colour profiles, we have 3 galaxies in common in $J-K$ 
and one in $R-K$, and here the agreement is better than 0.1 mag,
except near the center.

\subsubsection{Comparison with aperture photometry}

As a check on our photometric calibration we compare our data
to aperture photometry from the literature. The literature photometry
is from the compilation by de Vaucouleurs \& Longo (\cite{deV+Longo}).
We find aperture data for 5 of our galaxies.
On the average, our magnitudes are
0.145 mag brighter in $J$ and 0.061 in $K$. 
Table~\ref{TabApPhot} gives the differences 
(literature magnitudes minus ours) for the 5 galaxies. 

The differences are larger than expected from the
scatter in the standard stars, and consistent with the surface brightness
comparisons. They are, certainly in $K$, small enough that
they don't affect the final interpretation of the colours. 
The source of the discrepancies is unclear.
We have rerun the entire reduction process as a check, without applying
a linearity correction to the data, and obtained 
the same results. It should be noted that the effects of seeing 
on surface brightness profiles of highly-inclined galaxies are stronger
than on round galaxies. Also, 
the total number of aperture measurements is only 13 in each band. 
Both arguments together show that the discrepancy is minor, 
especially if
the errors in the literature surface photometry are underestimated.

\begin{table}
\caption{
	\label{TabApPhot}
	Comparison with aperture photometry}
\begin{center}
\begin{tabular}{lccc}
\hline
\hline
Galaxy & Nr. of & $\Delta$ J & $\Delta$ K \\
  &  measurements & & \\
(1) & (2) & (3) & (4) \\
\hline
NGC 5746 & 6 & 0.166 & 0.106 \\
NGC 5879 & 1 & 0.189 & 0.074 \\
NGC 5987 & 3 & 0.119 & -0.031 \\
NGC 7332 & 2 & 0.064 & 0.053 \\
NGC 7537 & 1 & 0.189 & 0.101 \\
\hline
{\bf Note:} $\Delta$=literature mean minus this paper.
\end{tabular}
\end{center}
\end{table}

\section{Discussion \& Conclusions}
\label{discussion}

In this data paper we present high-quality surface photometry for
a sample of early-type spiral galaxies in 6 bands, ranging from U (3700
${\rm \AA}$) to K (2.2 $\mu$m). The galaxies comprise a complete optically
selected sample of inclined spirals of type S0 -- Sbc. 
It is ideal to study the stellar populations of bulges,
since the bulges of early-type spirals are large, and in general
only one side of the bulges is obscured by the 
disk, so that the other is relatively free of extinction by dust and
of contamination by the disk. The infrared data have been obtained
with a 256 $\times$ 256 InSb array with a field of $\sim$ 80 $\times$ 80
arcsec, more than enough to contain the whole bulge, and the inner
galaxy disk. The effective seeing is approximately 1$''$ in the infrared
and 1.5 $''$ in the optical. 
Our photometry has been extensively compared with the literature
in BP94 (for the optical) and in this paper (for the infrared). 
For the surface brightness profiles the agreement is good. The absolute
calibration is accurate to 0.10 mag in $U$ and $B$, 
0.05 in $R$ and $I$, and 0.10 mag in $J$ and $K$. 
We provide surface brightness and colour profiles for the individual galaxies,
as well as measurements of absolute magnitudes, bulge-to-disk ratios,
bulge effective radii, disk scale-lengths, mean colours of bulges and disks, 
colour gradients of bulges and disks, and center shifts in each passbands.  

The sample may serve
as a standard sample for the colours of galactic bulges, replacing
the work of Persson \etal (1979) and Frogel \etal (1978).
It is complementary to the sample of de Jong \& van der Kruit 
(\cite{deJ+vdK}) in inclination, and to the sample of Peletier \etal (\cite{P++94})
in galaxy types. It has similar selection criteria as the sample
of Terndrup \etal (\cite{Terndrup++}). 

This is the first study of a moderately large number
of galactic bulges based on NIR data using 256 $\times$ 256
arrays. Compared to the recent study of Terndrup \etal (\cite{Terndrup++}), based on
58 $\times$ 62 InSb array data, 
our spatial sampling in the center is higher,
the photometry goes out further, and we cover more passbands 
(we also give $U$, $B$ and $I$).
Colours are treated in more detail in our paper.
In particular, we present 
minor axis colour profiles, which allow for the study of bulge
colours unaffected by extinction. 

We wish to use the electronic format of {\it New Astronomy} to present all the
calibrated images and colour maps electronically
for easy access and further processing by the community.  
We are confident that this will contribute to the furthering
of our understanding of the structure and formation history 
of galaxy bulges.  

\section*{Acknowledgements}

This article is based on observations taken at the United Kingdom
Infrared Telescope in Hawaii, operated by the Royal Observatory Edinburgh
on behalf of PPARC,
and from the Isaac Newton Telescope, operated by the RGO, on 
behalf of PPARC, at the 
Observatorio del Roque de los Muchachos of the Instituto de Astrof\'\i sica 
de Canarias. This research has made use of the NASA/IPAC Extragalactic Database (NED)   
which is operated by the Jet Propulsion Laboratory, California Institute   
of Technology, under contract with the National Aeronautics and Space      
Administration. The authors thank A. Vazdekis and M. Stiavelli for help
with the observations and A. Akalin for help with preparing the 
electronic manuscript.

\section*{Appendix A}

In this appendix we present three sets of figures, and four sets of tables,
which together form the whole available database on
the surface photometry of the inner parts of these galaxies.

\vfill\eject
\begin{figure}[ht]
\caption{
	\label{Monochromatic}
	Broad-band images of all the galaxies, in $U$, $B$, $R$, $I$,
$J$ and $K$, in GIF or FITS format.  Images are flatfielded,
calibrated, scaled and registered.  To download, click on the word 'FITS' 
and save to a local file.  
}
\end{figure}

\begin{figure}[ht]
\caption{
	\label{ColImages}
	$U-R$ and $R-K$ colour index maps for all the galaxies, in GIF format.
True-colour $URK$ composite images. See text.}
\end{figure}

\begin{figure}[ht]
\caption{
	\label{FigColProf}
	Colour index profiles for each galaxy, 
along the two semi-minor axes and along two axes 15\deg\ away from 
the major axis on the less-dusty side of each galaxy.  
The vertical dotted line is placed at 1 seeing FWHM.  
The two vertical solid lines indicate the limits of the range
used for the measurement of the colour gradients.  
}
\end{figure}

\vfill\eject

Figure~\ref{Monochromatic} contains FITS-format copies of the 
flatfielded, calibrated, scaled and registered
broad-band images of the galaxies 
in $U$, $B$, $R$, $I$, $J$ and $K$.  
Magnitudes corresponding to 1 ADU in each passband are given 
in the FITS header keyword PHOTCON.  
Pixel scale is 0.549 arcsec/pixel.  

Figure~\ref{ColImages} contains GIF format copies 
of the $U-R$ and $R-K$ colour index maps,
and a GIF format "true-colour" URK composite.  
In the colour index images, blue-red-yellow correspond 
to increasingly higher values of the colour index.  
The colour scale has been independently adjusted in each image 
to highlight the morphological components in the colour index maps.  
The true colour URK images are built using photometrically-calibrated
$K$, $R$, and $U$ frames as RGB colours. 
The colour scheme is the same for all the galaxies, ie. 
colour index comparisons among galaxies can be performed directly from 
the colours seen in these images. 
We have published 
a composite of the true-colour images of the entire sample in PB96a. 

Colour profiles along the minor and (near)-major axes are plotted in 
Figure~\ref{FigColProf}.  
The derivation of the profiles is described in  \S~\ref{BulCol}
for the bulges and \S~\ref{DisCol} for the disks. 
In each panel, each set of symbols corresponds to a different 
wedge aperture.  The vertical dotted line is drawn at 1 seeing FWHM.
The two vertical solid lines indicate the inner and outer limits
of the range used for measuring colour gradients. 

Table~7 gives the $R$-band surface brightness 
and the colour profiles 
for all the galaxies in the sample,
along the semi-minor axis where the dust effects are less pronouced
('dustfree' side).
The first column gives the radius in arcsec.  
Subsequent columns are in magnitudes per square arcsec.  
Errors, given in the $\pm$ columns, are in the same units. 
The header of each table gives the NGC number of the galaxy 
and the position angle of the semi-minor axis. 

Surface brightness and colour profiles 
along the other, more dusty semi-minor axis of each 
galaxy are given in Table~8.
The headers of these tables give the position angle, and 
a "D" denoting 'dusty side'.

Table~9 gives the $R$-band near-major-axis surface brightness 
and colour profiles for all the galaxies.
The first column gives the radius in arcsec along an axis
15\deg\ away from the disk major axis.
Subsequent columns are in magnitudes per square arcsec and represent
the surface  brightness azimuthally averaged in the two wedges on
the 'dustfree' side. The error in the next column is half the 
difference between the two wedges.

Table~10 gives the elliptically-averaged surface brightness
in $K$ as a function of major axis radius, 
together with the ellipticity (1-b/a), 
the position axis of the major axis, and the
4th order Fourier terms s4 and c4 (Carter 1980). In the outer parts
the ellipse shape has been kept constant, so that the surface brightness
is less affected by errors in those parameters. These parameters have
been fitted on the original $K$-band images, with a pixelsize of
0.291'', 
contrary to tables 7, 8 and 9,
which derive from NIR data rebinned to the larger, 0.549'' pixel size of
the optical data.

\end{document}